# Probing topological quantum matter with scanning tunnelling microscopy

Jia-Xin Yin[1]†, Shuheng H. Pan[2,3,4,5], M Zahid Hasan[1,6,7]†

[1]Laboratory for Topological Quantum Matter and Advanced Spectroscopy (B7), Department of Physics, Princeton University, Princeton, New Jersey 08544, USA.

[2]Institute of Physics, Chinese Academy of Sciences, Beijing 100190, China.

[3]School of Physics, University of Chinese Academy of Sciences, Beijing 100190, China.

[4]CAS Center for Excellence in Topological Quantum Computation, University of Chinese Academy of Sciences, Beijing 100190, China.

[5]Songshang Lake Material Laboratory, Dongguan, Guangdong 523808, China.

[6]Princeton Institute for the Science and Technology of Materials, Princeton University, Princeton, New Jersey 08544, USA.

[7]Materials Sciences Division, Lawrence Berkeley National Laboratory, Berkeley, California 94720, USA.

†Corresponding author, E-mail: jiaxiny@princeton.edu; mzhasan@princeton.edu

**The search for topological phases of matter is evolving towards strongly interacting systems, including magnets and superconductors, where exotic effects emerge from the quantum-level interplay between geometry, correlation and topology. Over the past decade or so, scanning tunnelling microscopy has become a powerful tool to probe and discover emergent topological matter, because of its unprecedented spatial resolution, high-precision electronic detection and magnetic tunability. Scanning tunnelling microscopy can be used to probe various topological phenomena, as well as complement results from other techniques. We discuss some of these proof-of-principle methodologies applied to probe topology, with particular attention to studies performed under a tunable vector magnetic field, which is a relatively new direction of recent focus. We then project the future possibilities for atomic-resolution tunnelling methods in providing new insights into topological matter.**

**Key points**

- Explorations of quasi-particle interference and Landau quantization behaviours in materials can be used to elucidate the nature of chiral or helical fermions and their interplay with symmetry-breaking order, such as magnetism, charge order or superconductivity.
- Bulk-boundary connectivity and its relationship with atomically resolved lattice and magnetic structure probed via topographic measurements are central to understanding emergent topology, such as in kagome lattices.
- Tunable vector magnetic field capability allows for exploring field-induced novel states in topological matter, including emergent magnetism, superconductivity and strongly correlated phases.
- In the presence of magnetic field control, spectro-microscopy plays an important role in identifying the nature of localized zero modes or in-gap states, such as the Majorana mode.
- Scanning tunnelling microscopy-based study on topological materials can be further extended for visualizing Wannier–Bloch duality and can be employed in sub-nanoscale engineering in developing quantum information science platforms.

## Introduction

The 1980 discovery of the quantum Hall effect observed in ultra-low temperature and strong magnetic field introduced topology in condensed matter physics[1,2,3]. However, over the last decade or so, topological electronic phenomena have also emerged in seemingly ordinary bulk materials under non-extreme conditions, leading to a new revolution in understanding the ubiquitous topological structure of quantum matter. These materials often host relativistic fermions, feature symmetry protected bulk-boundary correspondence, and exhibit quantized electronic excitations[1-11]. These topological fermions usually take the form of Dirac, Weyl, or Majorana solutions of the quantum field theory in high-energy physics. The topological bulk-boundary correspondence allows for making connections among various experimental techniques specifically sensitive to boundary or bulk or both. In addition, topologically protected quantized effects hold promise for potentially new technology, which may impact areas such as energy efficiency and quantum information science. The study of weakly interacting topological insulators featuring an insulating bulk and conducting boundaries, for instance, has achieved success, further driving the field to new frontiers to search for more exotic phases of matter[4,5]. With the advancement of the field, consideration of correlated materials is particularly fascinating, because they not only bring in the phenomena of magnetism and superconductivity at play[12-15], but also hold the potential to realize many-body entangled version of topological order[9-11]. More importantly, electron correlations present the possibility of emergence and offer the potential for unexpected discovery.

The topological materials with strong correlations can exhibit spin, orbital ordering or superconducting instabilities with intrinsic magnetic, orbital, or electronic anisotropy[9-11], which call for advanced experimental tools to probe and elucidate the interplay between correlation and topology. Notably, external magnetic fields that break time-reversal symmetry can act as a strong perturbation to magnetism and superconductivity, as well as modify the band structure in the presence of strong spin-orbit interaction[16]. The magnetic field applied on an electronic system leads to a nontrivial topology: the magnetic flux quantum ($h/e$) and quantum Hall conductance ($Ne^2/h$, related to Chern number $N$, a topological invariant) are governed by the same set of fundamental constants[17] including the Planck's constant $h$ and elemental charge $e$; the vector nature of the field can differentially interact with the chirality of topological matter to provide access to effects related with the topological invariant[11]. While transport techniques under a vector magnetic field can play a vital role[18], conventional transport for bulk materials (that are often not gate tunable) is only sensitive to the electronic properties at the Fermi level, limiting its exploration necessary for probing the band interconnectivity and topology. Traditionally, in comparison with optical, X-ray, and neutron techniques, photoemission spectroscopy is regarded as a powerful technique to map the interconnectivity and topology of the electronic band structure[19]. However, an external magnetic field can be detrimental to detecting photoelectrons requiring complex orbital corrections, severely hindering its field application. In this regard, STM is an indispensable high-resolution spectroscopic technique that can work under a tunable vector magnetic field. In this review, we start out with the basics of STM and elaborate on its application in exploring electronic and magnetic properties of topological materials. We elaborate on several proof-of-principle methods to link STM signals with quantum topology, and broadly discuss how the technique can supplement, bridge, and complement the traditional photoemission and transport, as well as other related methods in this research area. Finally, we point out several unique directions to extend the STM technique and its applications in exploring emergent topological matter.

## State-of-the-art STM technique

Low-temperature STM is a cutting-edge microscopic and spectroscopic technique that can be employed under a strong magnetic field with energy ($E$), space ($\boldsymbol{r}$), and scattering-vector ($\boldsymbol{q}$) resolution[20-28]. It utilizes the quantum tunneling principle[29-31] to probe the surface morphology and the local density of states (LDOS) of materials with atomic-scale precision and (sub)meV energy resolution. When the scanning tip and sample are atomically close, their wavefunctions overlap and electrons can tunnel through their vacuum gap. The fast decay of the wavefunction leads to the exponential sensitivity of the tunnelling current $I$ to the sample-tip distance $d$. This exponential dependence of the electronic signal on physical distance allows for accurate feedback control of the scanning motion of STM tip to probe the surface morphology, with the atomic spatial resolution. Furthermore, fixing $d$ by a rigid STM mechanical structure, through sweeping the bias voltage V added between the sample and the tip, the energy range of the tunneling states associated with both the sample and the tip[31] can be precisely tuned (Fig. 1**a**). By selecting the tip material so that its DOS is energy independent and assuming that tunnelling matrix element is space independent, the differential tunnelling conductance is proportional to the LDOS($E, \boldsymbol{r}$) of the sample at the location of the tip, convoluted with the energy derivative of the Fermi-Dirac distribution $f'(E,T)$ at the measuring temperature $T$: $\frac{dI}{dV}(E, \boldsymbol{r}) \propto \int \text{LDOS}(E - \omega, \boldsymbol{r}) f'(\omega) d\omega$. The mathematical details for its derivation can be found in Ref.[21]. Essentially, the dI/dV signal samples the LDOS($E$, **r**) with a thermal-induced finite energy resolution of $3.5k_BT$, where $k_B$ is the Boltzmann constant[20-28]. We also note that scanning tunnelling spectroscopy becomes more challenging by progressively moving away from the Fermi, where a non-constant tip DOS as well as the larger background related to the increased tunneling probability can complicate the interpretation of the spectroscopic data.

The combination of scanning and tunnelling provides comprehensive spectro-microscopy of material, including spontaneous morphology visualization and spectroscopic imaging. Figure. 1**b** illustrates the microscopy of a material (for example, CoSn, REF.[32]). The topographic image here reveals the surface morphology, which resolves both the $Sn_2$ honeycomb lattice layer and the $Co_3Sn$ kagome layer with atomic resolution. This atomic lattice resolving capability is crucial in probing the respective electronic structure properties of different lattice geometries, whose underlying real-space electronic structure can be simultaneously obtained by the dI/dV mapping. An energy slice of the dI/dV mapping can demonstrate the electronic difference between two atomic layers (lower left panel of Fig. 1**b**). In particular, the defect-induced oscillation pattern is more evident on the kagome layer, and a Fourier transform of the dI/dV map taken on the kagome layer provides the essential quasi-particle interference (QPI) signal in this kagome paramagnet (lower right panel of Fig. 1**b**). The topography, dI/dV spectrum, and associated QPI constitute the primary data structure types for typical STM measurements.

To facilitate high-resolution microscopy, a rigid STM design is required, capable of an ultra-stable picometer-level control of the tip position on the sample despite all potential environmental sources of noise. As an example, the STM shown in Fig. 1**c**, which utilizes the cooperative motion of multiple tightly clamped piezo legs against the moving scanner in a triangular configuration[33-35], is a useful rigid design widely adapted for applications at low temperature (down to 10mK) and under a tunable (vector) magnetic field[36-45]. In these systems, a superconducting solenoid is used to generate a strong magnetic field along the $z$-axis, providing a primary vector field (commercially up to ±18T). Pairs of split superconducting coils can be further included in the system to generate magnetic fields in the horizontal directions (commercially up to 5T), providing a 3D vector magnetic field. Moreover, in contrast to magneto-transport measurements under a ramping magnetic field, STM measurements are usually performed under a stabilized magnet field to minimize the mechanical drift and electrical noise, which is associated with delicate STM tip retractions and approaches between ramping to different fields. Therefore, a full set of systematic STM data under a vector magnetic field requires a combination of reliable instrumentation, skillful control of the tip, and suitable material platforms.

In the STM research of topological materials, besides the aforementioned technical challenges, a major question is how to link the tunnelling signal with the underlying topology. In this regard, we point out a general guideline. For a spectroscopic technique, new information often manifests as special spectroscopic patterns or peaks. Firstly, we search for an anomaly by comparing the tunnelling signal (unusual patterns and peaks) with those on well-known topologically trivial materials, including elemental metals and 2D electron systems. Secondly, we perturb the anomaly by tuning as many parameters as possible to provide substantial constraints for its interpretation. Lastly, we refer to other complementary experimental or theoretical techniques (Table1, as an example) for a comprehensive understanding. More specifically, we highlight several proof-of-principle methodologies that have been developed in this field in the following discussion.

## Proof-of-principle methodologies

**Quasi-particle interference method**
Prior to its application in the study of topological materials, QPI signal was observed in noble metals and superconductors[46-51], and has been developed as a general STM methodology. The QPI is reflected in the spectroscopic pattern through the Fourier transform of the measured real-space dI/dV image. The QPI signal stems from Friedel oscillations (or standing waves) around defects detected in real-space measurements (Fig. 2**a**). The periodicity of such oscillations gives rise to the *q* wavevector that is associated with the elastic scattering of quasi-particles and their interference between two momentum states (*k*$_1$ and *k*$_2$).

Through measuring the QPI patterns as a function of energy, we can obtain the energy variation of *q*, which is the autocorrelation of the electronic band dispersion $E(k)$. For the same band dispersion, the intensity distribution of QPI pattern can be different, depending on the quasi-particle scattering geometry. This geometry is related to the charge, spin, and orbital texture of the electron distribution, which is associated with the spectral function $A(k, E)$ —imaginary part of the Green's function. Firstly, the spectral function intensity at different momenta can vary, and the QPI intensity is proportional to the intensities of the spectral function at *k*$_1$ and *k*$_2$. Secondly, the electronic structure can feature a momentum dependent spin texture, and the QPI intensity is substantially reduced if the spins at *k*$_1$ and *k*$_2$ are in opposite directions, especially for QPI assisted by nonmagnetic defects. Thirdly, the electronic structure can feature a momentum dependent orbital texture, and the QPI intensity is substantially reduced if the orbitals at *k*$_1$ and *k*$_2$ are orthogonal. In addition, there are also cases that QPI intensity can be geometrically enhanced at *q* that connects two parallel pieces of the band structure within a constant energy contour, known as the nesting effect[52]. These factors form the basic selection rules for the general interpretation of the QPI data, and the mathematical details for such a physical picture including the joint density of states (JDOS) approximation can be found in Refs. [24-27]. For simple isotropic systems, the QPI signal near the Fermi level is a nearly isotropic ring, reflecting a plain circular Fermi surface[49]. Thus, anisotropic, spot-like, or disconnected arc QPI data often presents a hint of an unconventional band structure that could be further related to nontrivial band topology. As QPI data convolutes the scattering effects from charge, spin, and orbital textures, its final interpretation would require accurate knowledge of the bare band structure and scattering processes. In practice, QPI patterns can be simulated by considering the band structure obtained from angle-resolved photoemission data or density functional calculations under different scattering geometry considerations (note that photoemission cannot directly probe scattering geometry), and then it can be compared to the measured QPI pattern to gain insight of topology.

Here we highlight several examples of QPI applications in topological materials. A rich QPI pattern is typically observed on topological insulator $Bi_{1-x}Sb_x$ (Fig. 2**b**, top panel, Ref.[53]). The QPI signal is beyond simple circular pockets, indicative of nontrivial scattering geometry. The QPI data is compared with angle-resolved photoemission spectra[54], where there exist three sets of surface Fermi pockets with spin-momentum locked chiral spin textures. The QPI data should be related with the quasi-particle scatterings among these pockets mediated by the random spatial potentials associated with the alloying. By applying the charge and spin selection rule, there can be three dominant scattering vectors along the Γ-M direction (Fig. 2**b**, white arrows in middle panel) while direct backscattering is forbidden. A QPI simulation based on this selection rule explains the observed QPI pattern along the Γ-M direction (Fig. 2**b,** bottom panel) as well as other directions. The QPI demonstration of the lack of backscattering, therefore, supports the chiral spin texture, which is associated with the Berry phase of the topological matter. A simpler QPI pattern is observed on topological insulator $Bi_2Te_3$ (Fig. 2**c**), where the deposited Ag impurities assist the QPI signal[55]. This material also shows a simpler surface state with only one hexagonal Fermi pocket[56-58] (Fig. 2**c**, bottom panel), thus is close to a simple 3D topological insulator. Remarkably, the anisotropy of the QPI pattern (stronger along Γ-M direction) is distinct to that of the Fermi surface intensity (stronger along Γ-K direction), providing evidence that direct backscattering is forbidden. The observed scattering vector along Γ-M direction can be understood as the effect of hexagonal warping on surface states with a chiral spin texture[59], which will not be detectable without warping as seen in $Bi_2Se_3$ and $Bi_{1.5}Sb_{0.5}Te_{1.7}Se_{1.3}$ (REF.[60,61]). The QPI signal in $Bi_2Te_3$ supports the unusual spin-momentum locking, which is a consequence of the underlying topology[4,5].

In topological insulators, the quantum topology is protected by the time-reversal symmetry, thus magnetic impurities that break time-reversal symmetry can lead to strong perturbations. In particular, magnetism can modify the helical spin-momentum locking near the Dirac cone, and the magnetic impurities can mediate scattering between spin-reversed states, both of which would create new scattering channels in the QPI that are otherwise forbidden. Systematic experimental efforts along this direction are focused on the magnetic doped $Bi_2Te_3$ materials[62-65], revealing new results. While bulk magnetization has been seen in $Bi_{2-x}Mn_xTe_3$ and $Cr_x(Bi_{0.1}Sb_{0.9})_{2-x}Te_3$, no apparent new QPI signal is detected along Γ-K direction[62,65], which seem to suggest the robustness of the forbidden scattering in these systems. However, the QPI of $Bi_{2-x}Fe_xTe_3$ and $Bi_2Te_3$ with surface Mn impurities reveal a new scattering channel along the Γ-K direction emerging around 200meV above the Dirac cone[63,64]. This new scattering vector is identified as evidence for time-reversal symmetry breaking. These experiments suggest that details of the magnetic order pattern associated with the magnetic impurities are crucial to observe such an interplay between scattering and topology.

Besides the topological insulator, QPI has also been applied to probe other topological materials, including topological crystalline insulators and Weyl semimetals[66-77]. In topological crystalline insulator (Pb,Sn)Se and Weyl semimetal NbP, the QPI anisotropy is linked to the orbital texture of the topological surface states[66,67]. In Weyl semimetal TaAs and $Mo_xW_{1-x}Te_2$, signatures of the Fermi arcs, as well as the surface-bulk connectivity, have been studied[68-77]. More recently, the QPI method is used to study the topological surface states in superconductors $PbTaSe_2$ and $β-PdBi_2$ (REF.[78,79]), uncover the tunable nematicity in topological kagome magnet $Fe_3Sn_2$ (REF.[80]), resolve the heavy Dirac fermion in Kondo insulator $SmB_6$ (REF.[81]), and detect fermion-boson interplay in topological kagome paramagnet CoSn (REF.[32]). We also note that while QPI can visualise topological scattering geometry, QPI alone is often not sufficient to prove nontrivial topology. For instance, backscattering between states related by time-reversal symmetry is also forbidden in Rashba systems[82], which may not involve with nontrivial topology. Moreover, in systems with complex fermiology, QPI patterns may emerge to be more complex, making it challenging to pinpoint the topology.

**Landau quantization method**
Magnetic field induced Landau level quantization is reflected in the spectroscopic peaks through measuring dI/dV signal. Landau levels are a sequence of discrete energy values formed due to the quantization of the cyclotron orbits of charged particles in magnetic fields. Prior to its application to topological materials, Landau levels in the LDOS have been observed in other materials, including 2D electron systems, graphite, and graphene[83-87]. It has been established that the Landau levels of the Dirac band are markedly different from those of a parabolic band, owing to the Berry phase effect. Figure 3**a** illustrates the Landau levels of a 2D Dirac band as a function of the magnetic field ($B$), which can be described by $E_n = E_D + \text{sgn}(n)v\sqrt{2|n|e\hbar B}$, $n$= -2, -1, 0, 1, 2, …, where $E_D$ is the energy of the Dirac cone relative to the Fermi level and $v$ is the velocity of the Dirac band dispersion. In contrast, the energy of Landau levels of a parabolic band is described by $E_n=\varepsilon_0 \pm (n+\gamma)e\hbar B/m^*$, $\varepsilon_0$ the energy of the band edge, $\gamma$ is the Onsager phase, and $m^*$ the effective mass of the band.

The unique features of Dirac Landau levels include the following. The energy of the zeroth Landau level does not change with $B$, the energy separations of Landau levels are not equal and decrease with increasing $n$ and the non-zeroth Landau levels display a quadratic dispersion with $B$. All of these features can be directly identified by the sequence of spectral peaks in the dI/dV data as a function of magnetic field strength. Moreover, the quantization nature of Landau levels enables a direct comparison with analytical models and magneto-transport data. The band dispersion can be extracted from Landau levels to compare with angle-resolved photoemission data and density functional theory. Therefore, Landau level imaging is a powerful method to diagnose the existence of Dirac or other topological fermions with linear band crossings in materials.

The detection of Landau level formation in real-space often requires materials to have few defects. A relevant length scale is the magnetic length, $l = \sqrt{\hbar/eB}$, which is associated with the wavelength of the zeroth Landau level. Thus, the STM detection of Landau levels at a few Tesla would imply that the average defect inter-distance is larger or of the same order as $l$, and the atomic defect rate is normally less than 0.1%, pointing to their quantum-limit nature. Here we highlight several examples for Landau level imaging in materials. The d$I$/d$V$ spectra taken at different magnetic fields in topological insulator $Bi_2Se_3$ show evidence of the Landau quantization of surface Dirac fermions (Fig. 3b)[88,60]. The Landau levels at different magnetic fields taken together provide a Landau fan diagram. The zeroth Landau level is identified as the peak at -200meV that barely shifts with $B$, and the Landau levels for $n > 0$ are subsequently identified which exhibit energy differences proportional to $\sqrt{nB}$ with respect to the zeroth Landau level. Fundamentally, the Berry phase of Dirac fermion would also imply that leaving the zeroth Landau level empty or completely filled gives rise to the half-integer quantum Hall effect. This expectation is visualised in the later transport experiment. The fixed zeroth Landau level relative to $B$ can also be seen in the gate-tuned transport Landau fan measurement of $BiSbTeSe_2$, which is a close cousin of $Bi_2Se_3$ and exhibits surface-dominating conduction (Fig 3c)[89]. When the bottom surface is gated through the zeroth Landau level, an odd integer quantum Hall plateau is observed. This observation is consistent with a half-integer quantum Hall effect of two degenerate Dirac bands arising from two surfaces[89]. With real-space imaging capability, researchers have also investigated the perturbations of local atomic impurities or charge potentials to the Landau levels[88,90-93]. Intriguingly, although the surface Dirac band exhibits spin-momentum locking in $Bi_2Se_3$, its Landau levels can be shifted and split by a charge potential in real space (Fig. 3**d**), indicating their two-component nature with the possible formation of a real-space spin texture[91].

Besides the topological insulators, Landau level imaging has also been applied to study other topological materials, including topological crystalline insulator (Pb,Sn)Se (REF.[94,95]), Dirac semimetal $Cd_3As_2$ (REF.[96]), as well as Chern magnet $TbMn_6Sn_6$ (REF.[97]). In these materials, the Landau levels exhibit additional complexity beyond features for a single Dirac cone. The complexity is associated with the interplay between symmetry-breaking order and band topology, which can often be described by the analytical Landau level model. The Landau fan of (Pb,Sn)Se features three sets of zeroth modes that do not change energy with varying $B$ fields, indicating the coexistence of massless and massive Dirac fermions (Fig. 3**e**). The mass acquisition associated with the observed lattice distortion is a consequence of the crystalline quantum topology[94]. Through further chemical engineering, the Landau level imaging method helps to identify a topological quantum phase transition associated with the Dirac mass generation[95]. The Dirac-like Landau fan is obtained in Dirac semimetal $Cd_3As_2$ (Fig. 3**f**), where the Landau levels can be attributed to bulk Dirac fermions projected on surface[96]. Two sets of Dirac-like Landau levels are observed, which provides an early indication of the splitting of Dirac fermions into Weyl fermions by the applied field.

**Emerging methods to discover topology**

QPI and Landau level quantization are both effective methods to characterize the nature of topological fermions. However, these STM studies in topological materials are often guided by other techniques and predictions that present the early evidence for topology. With the development of newer STM methodologies in probing topological matter, a question arises: can STM play a leading role in discovering new topological materials/phenomena? To achieve such innovation, we need to have a profound understanding of the fundamental principles and the measurable consequences of quantum topology as well as material science at play. In this section, we outline case studies showing emerging methods where STM has played a leading role in unveiling the topological nature of a material.

**Topological correspondence**

The topological correspondence principle, which emphasizes the topology governed connection between different physical properties, is a powerful way to establish quantum topology in STM studies. The bulk-boundary connectivity, for instance, is one topological correspondence that has been pointed out since the early research on the quantum Hall effect[98]. In materials, the nontrivial topology of the bulk electronic structure can induce robust edge states around the boundary. As STM has the advantage of resolving step edges, the detection of robust edge states can often serve as an indication for nontrivial bulk topology, with examples including the topological step edge states observed in Bismuth, $Bi_{14}Rh_3I_9$, $ZrTe_5$, $HfTe_5$, $FeSe/SrTiO_3$, (Pb,Sn)Se, Bismuthene, $WTe_2$, $WSe_2$, and $TbMn_6Sn_6$ (REF.[99-111,97]). This method is particularly powerful when the bulk energy gap can be independently identified away from the edge, as in the case of $ZrTe_5$, $HfTe_5$, $WTe_2$ monolayer, Bismuthene, and $TbMn_6Sn_6$. Unusual fermions arising from lattices with unique geometry (such as honeycomb, kagome, Lieb, and chiral lattices) are another relevant example of topological correspondence, as topologically nontrivial band degeneracies and band singularities can be protected by special lattice geometry. This correspondence is particularly relevant for STM, since STM has the advantage to resolve different lattice layers in a complex material. A well-known example is graphene, and a recent example includes the topological kagome magnet family (Box. 1) where STM can play a leading role in many novel observations[80,97,112-121,32].

Here we take the STM study[97] on kagome Chern magnet $TbMn_6Sn_6$ as a case in point to elaborate on the topological correspondence (Fig. 4**a**). The kagome lattice naturally hosts Dirac electrons at the Brillouin zone boundaries. The inclusion of spin-orbit coupling and out-of-plane ferromagnetic ordering in the kagome lattice effectively realises the spinless Haldane model by generating Chern gapped topological fermions[122-124]. Essentially, the spin-polarized Dirac fermions in momentum space will open an energy gap with a nonzero Chern number (Chern gap), whereas the magnetic kagome lattice carries a chiral edge state correspondingly (Fig. 4**a**). Among many other kagome magnets, $TbMn_6Sn_6$ uniquely features a pristine Mn kagome lattice with strong out-of-plane magnetization. STM study further finds that the Mn kagome lattice is almost free from defects and exhibits unique Landau quantization, while another hexagonal lattice in the same material does not feature Landau quantization. The Landau fan analysis demonstrates a spin-polarized Dirac band with a large Chern gap ($\Delta$) (Fig. 4**b**). The Landau levels can be described by $E_n = E_D \pm \sqrt{(\Delta/2)^2 + 2|n|e\hbar v^2 B} - \frac{1}{2}g\mu_B B$, with $E_D$ = 130 meV, $\Delta$ = 34 meV, $v$ = 4.2 ×10$^5$ m/s and Landé *g*-factor, $g$ = 52. Based on these parameters extracted from the Landau levels, the band dispersion can be obtained, which shows agreements with occupied bands observed by angle-resolved photoemission data and the density functional calculations. Precisely at the Chern gap energy position, a pronounced step edge state is observed (Fig. 4**c**), and the quasi-particle scattering is substantially reduced as measured by the QPI on the sample edge, both of which are consistent with the existence of a chiral edge state within the Chern gap. Based on the obtained Dirac energy and Chern gap size, the Berry curvature contribution[18] to the anomalous Hall conductivity can be estimated as $\sigma_{xy} = \frac{\Delta}{2E_D} * e^2/h = 0.13\ e^2/h$ per kagome layer, which shows an agreement with the intrinsic anomalous Hall conductivity ($\sigma_{xy}^{int}$) obtained by magneto-transport as $\sigma_{xy}^{int} = 0.14\ e^2/h$ per kagome layer within the error bar of the measurement (Fig. 4**d**). The agreements between Landau levels imaging, edge state (QPI) imaging, and anomalous Hall conductivity collectively provide evidence for the topological bulk-boundary-Berry correspondence in identifying this Chern magnet. This case study also indicates that with the guidance of topological correspondence principles, STM studies can present original evidence for topological matter.

**Vector magnetic field control.**
Besides the topological correspondence principles, the vector magnetic field capability of the STM technique provides a new channel to probe and manipulate topological matter. For instance, this technique has been used to study vortices in anisotropic superconductor NbSe$_2$ (REF.[125]). Recently, it has also been applied to probe the vortices and QPI in superconductors Cu$_x$Bi$_2$Se$_3$, LiFeAs and Bi$_2$Te$_3$-NbSe$_2$, all of which have topological surface states[126-130]. The interplay between anisotropic Cooper pairing and topological surface states can be studied in detail. Another emerging direction is the application in topological magnets, where the vector magnetic field can control topological band structure owing to the strong spin-orbit coupling and magnetic exchange interaction in these materials.

A notable example is the STM study of topological soft magnet Fe$_3$Sn$_2$ (REF.[80]), in which the magnetization direction can be effectively controlled through a vector magnetic field. STM study of this material reveals that upon increasing a magnetic field along the *c*-axis, a quantum state exhibits a progressive energy shift, which is in strong correlation with the bulk magnetization curve but is in stark contrast with the conventional Zeeman effect. Moreover, the rotation of the in-plane magnetization direction introduces an energy shift with strong two-fold anisotropy concerning the rotation angle. This nematicity is consistently observed through magneto-transport with rotating an in-plane field[131]. The vector magnetization induced energy shift exhibits a node along the crystalline *a*-axis (Fig 5a, top panel), indicating a spontaneous magnetization along the *a*-axis. Through QPI dispersion study, this energy shift is associated with the magnetization tuning of the Dirac gap (Fig. 5**a**, bottom panel) with the quantum state corresponding to the band bottom of the upper Dirac branch. Intriguingly, the d*I*/d*V* mapping of this quantum state reveals spontaneous QPI nematicity along the *a*-axis (Fig. 5**b**, top panel). As this QPI data is taken in absence of the applied magnetic field, the observed nematicity direction supports spontaneous magnetization along the *a*-axis, and demonstrates an internal consistency with the anisotropic energy shift data in Fig. 5**a**. Magnetization along other directions by applied the magnetic field can alter, and thus control, the electron scattering symmetry (Fig. 5**b**). Such vector magnetic field control of the scattering geometry can be explained in the context of the interplay between vector magnetization and the underlying spin texture of the Dirac fermions[80].

Another example is the STM study of topological hard magnet $Co_3Sn_2S_2$ (REF.[113]), in which Berry phase induced orbital magnetism and flat band physics can be resolved[132-134]. STM study of this material reveals that a flat band contributes to the tunnelling signal and has a highly unusual vector magnetic field response. The flat band state (associated with the blue color in Fig. 5**c**) exhibits a progressive energy shift with the magnetic field, and it shifts to the positive energy direction irrespective of the vector magnetic field direction along the *c*-axis. The shift rate can be used to calculate an effective magnetic moment of $-3\mu_B$. This field-induced energy shift of the flat band state is highly unusual from two aspects. Firstly, the field direction independent energy shift is in strong contrast with the conventional Zeeman effect. As the magnetic field flips the bulk magnetization direction, it introduces a magnetization polarized Zeeman shift. Secondly, the negative magnetic moment is beyond the spin Zeeman effects ($\sim+1\mu_B$, as the flat band features spin-up in density functional theory) and indicates the dominant negatively polarized orbital magnetization. Indeed, the Berry phase induced orbital magnetism for the flat band calculated from density functional theory (Fig. 5**d**) shows the same sign and order of magnitude as that obtained by STM. The STM experiment thus unveils the topological effect of the magnetic flat band state. More recently, STM studies under a vector magnetic field show that the impurity resonances also exhibit negative orbital magnetism[120,121], implying ubiquitous spin-orbit coupling and quantum phase effects in the topological magnet $Co_3Sn_2S_2$.

**Localized topological zero modes**.
Similar to its vector magnetic field capability, STM is a unique tool to search for exotic quantum states at the extreme local scale in a material, where ***k*** is no longer a suitable quantum number. These local quantum states include the single-atomic impurity induced resonance[36] and the vortex core states[135] in superconductors. The combination of magnetism, topology, and superconductivity enables the search for Majorana fermions[136-145]. Majorana modes can manifest as a robust zero-energy peak (topological zero modes) inside a superconducting gap in the tunnelling spectra. Systematic efforts have been made to design artificially hybrid structures to visualise the topological zero modes based on various theoretical proposals; this is extensively discussed in other review articles[139-145]. Here we highlight an alternative approach for STM to search for originally unpredicted and naturally occurring topological zero modes in bulk Fe(Te,Se) crystal and related materials, and this approach can be another useful guideline for future experimental discoveries.

Before the topological nature of Fe(Te,Se) gained appreciation, a robust zero-energy bound state at the native interstitial Fe impurity[38] was observed by an early STM work (Fig. 6**a**). While measuring slightly away from the impurity, the bound state fades away but remains centred at zero-energy. Remarkably, this state is robust against even an 8T applied magnetic field, which would, otherwise, have induced a large splitting for any spin degenerate state. This observation has no explanation in terms of classical impurity states in superconductors with *s*-wave symmetry. However, this state bears many characteristics of the Majorana mode proposed for topological superconductors, suggesting possible nontrivial topology in the system[38]. This observation stimulated further density functional theory calculations of Fe(Te,Se) which demonstrated nontrivial band topology[146-148], also confirmed by angle-resolved photoemission study[149]. The discovery of nontrivial quantum topology in these materials opened up new research directions in the study of iron-based superconductivity.

The observation of a robust zero-energy bound state at the Fe impurity stimulated further advanced theoretical proposals on topological excitations, including the concept of the quantum anomalous vortex, formally in analogy with the quantum anomalous Hall effect[152]. In *s*-wave superconductors with strong spin-orbit coupling, magnetic impurity ions can generate topological vortices in the absence of external magnetic fields. Such vortices, dubbed quantum anomalous vortices, support robust Majorana zero modes when superconductivity is induced in the topological surface states. The existence of quantum anomalous vortices at native interstitial Fe impurities is also consistent with the spontaneous Nernst signal in thermal transport measurement[160]. The quantum anomalous vortex appears to be a more robust way to generate Majorana zero modes than the field-induced vortex method[152], as the topologically trivial vortex state is pushed to higher energies. This argument is consistent with the STM experiments[129,156] in LiFeAs. In this system, the field-induced vortex shows no zero-energy bound state even with systematic Co doping, which is attributed to the contamination of topologically trivial vortex states[129]. On the other hand, depositing Fe adatoms can generate a zero-energy bound state (Fig. 6**b**), which is robust against vector magnetic field perturbations[156]. These findings suggest that magnetic adatoms on superconductors with topological surface states can be a field-free platform for exploring topological zero modes. The robust zero-energy bound state is also observed on the Fe adatom deposited on Fe(Te,Se), which is insensitive to the magnetic field and can be further tuned through interaction with the STM tip[156].

Another unexpected observation is on the native line-defect in Fe(Te,Se)/$SrTiO_3$ (Fig. 6**c**), where the zero-energy peaks emerge at the line ends[159]. Benefiting from a higher superconducting transition temperature from interface coupling, this can serve as a high-temperature platform for topological zero modes. Another native defect, the crystalline domain wall in Fe(Te,Se), also exhibits nontrivial excitations[157]. The flat in-gap state is consistent with a linearly dispersing mode in 1D, suggesting the possibility of a propagating topological fermion.

With the identification of quantum topology in Fe(Te,Se), a natural direction is to search for the Majorana zero mode in magnetic field induced vortices, based on the original theoretical proposal[137,138]. This direction has been fruitful. STM measurements have found that there exist vortices with pronounced nonsplitting zero-energy bound state[150] as evidence for a topological zero mode (Fig. 6**d**). Ultra-high energy resolution STM further confirms their zero-energy nature for the core states of some vortices (Fig. 6**e**), and reveals ratios of zero mode-carrying vortices as a function of magnetic field[153]. The topological nature of the vortices with and without zero-energy modes has been further explored[154,158] by an inspection of the side peaks of their core states and examination of the tunnelling conductance quantization[162]. In parallel, the non-splitting zero-energy vortex bound state was observed in another iron-based superconductor (Li,Fe)OHFeSe (Fig. 6**f**) to exhibit nearly quantized conductance[151,155]. The nontrivial native defects and magnetic-field induced vortices collectively present a zoo of potential Majorana platforms in iron-based superconductors. These experiments also stimulate discussions on the next frontier of Majorana research, including the possibility of braiding operations to demonstrate the non-abelian anyonic statistics[136], which will be the eventual smoking-gun evidence for Majorana interpretation of the existing results.

## More to discover

STM is becoming an advanced technique in probing and uncovering emergent quantum effects and topological phases. Looking forward, there remains substantial room to improve the capabilities of STM. On the instrumentation side, for example, a new direction is the ability to rotate the STM head in a simple high field solenoid, which can push the effective vector magnetic field magnitude far beyond the current technical limit. This is possible by extending the STM design to incorporate rotational functionality. On the technique side, it has been possible to perform experiments under a ramping mode while compensating for the tip drift when the field ramping rate is sufficiently low[97]. This methodology allows for the faster Landau fan imaging, for example, in the topological magnet $TbMn_6Sn_6$, and will be valuable in the application to field-rotation coupled STM measurements. Lastly, it has been recently pointed out that an electric field-induced change in the electrons' charge density orientation can be an effective way to probe Berry curvature field[163]. Therefore, the implementation of vector electric field tuning in STM systems will likely be another fruitful direction to extend the reach of the technique to access exotic topological phenomena.

The tunable vector magnetic field capability coupled with STM promises great potential in visualising quantum engineering routines applied on topological materials. Besides the topological kagome magnet family with various magnetic structures, there are a number of other correlated topological materials that feature novel magnetic response which can be explored using STM. For instance, singular angular magnetoresistance has been observed in a magnetic nodal semimetal CeAlGe (REF.[164]) and magnetic nodal-lines have been experimentally observed in a soft magnet $Co_2MnGa$ (REF.[165]), both of which are excellent candidates for study of the interplay between Weyl-line fermions and vector magnetization which is possible by STM. It is predicted that $EuIn_2As_2$ can either be an axion insulator or a topological crystalline insulator (exhibiting different hinge states) depending on the direction of the magnetic moment[166]. This promises to be another fertile platform for STM under a vector magnetic field to explore, manipulate and control the associated high-order topology[167]. Moreover, nearly ferromagnetic spin-triplet superconductivity has been detected in $UTe_2$ (REF.[168]) and chiral edge states were observed by low-temperature STM (REF.[169]). Thus, low-temperature STM under a vector magnetic field can be used to elucidate the interplay between magnetism and chiral superconductivity. It has long been known that an in-plane magnetic field can induce chain vortex matter in $Bi_2Sr_2CaCu_2O_{8+\delta}$, while a strong spin-orbit effect was recently observed in the same material[170]. STM with an in-plane magnetic field can explore the spectral characteristics of the chain vortex matter and the possibility for high-temperature 1D topological superconductivity in cuprates. There are also possibilities to explore more exotic systems, including a system of Majorana zero modes with random infinite-range interactions—the Sachdev-Ye-Kitaev model that theoretically exhibits an intriguing analog to the horizons of extremal black holes[171,172]. Materials including Fe(Te,Se) with random excess Fe impurities carrying zero modes may be explored by STM to test these speculative but tantalizing ideas.

One important class of topological correspondence is the Wannier–Bloch duality[173,174]. Wannier functions are the localized molecular orbitals, and Bloch wavefunctions are related to the electron energy levels in a periodic crystal. The theoretical realisation of Wannier–Bloch duality links the local chemistry and extended wavefunction property approaches to describe electronic states, leading to a modern understanding of electric polarization and orbital magnetization[173], as well as allowing systematic first-principles predictions of topological materials[174]. STM can be used to locally probe molecular orbitals[116], which can additionally be sensitive to Berry phase induced orbital magnetism with magnetic field perturbations[113]. The ability to visualise Wannier–Bloch duality correspondence by STM can lead to new experimental directions in elucidating topology.

Quantum information science has emerged as one of the major research frontiers[175], which includes STM-based characterization and fabrication techniques to enable the bottom-up construction of qubits from the atomic components[176,177,120]. Particularly notable is the progress in visualising topological materials at the atomic scale that could yield inherently error-protected qubits expected to feature a high level of digital fidelity. However, a clear demonstration of topological protection of quantum information in materials remains an open question for further exploration. Collaboration between STM and device engineering would further enable the development of models for optimizing topological materials selection for desired functionality based upon controllable properties, such as density of states, tunnelling energies, and the qubit-relevant characterization of materials-related decoherence channels.

Besides these exciting frontiers, there are potential research areas where the related questions are less defined and remain open-ended, including whether STM technique can be utilized to explore the many-body entangled version of topological order[9-11]. With the quantum level interplay of geometry, correlation, and topology resolved at the atomic scale, it is conceivable that there are more important yet hitherto unknown emergent phenomena to be discovered through STM research on topological matter. In his last public talk in 2019, P. W. Anderson was asked: "What is your best suggestion to young researchers when they encounter a new phenomenon?" Anderson responded: "Patience, and it can take a lifetime long (to understand certain emergent phenomena substantially)." His remark highlighted the naturality to encounter "unknown unknowns" in quantum materials. We believe that time is ripe in the field where tunnelling into topological matter will likely lead to genuinely new phenomenon in the next few years, which will reap multiple lifetimes' worth of rewarding scientific vistas for our efforts now.

**Figures and Captions**

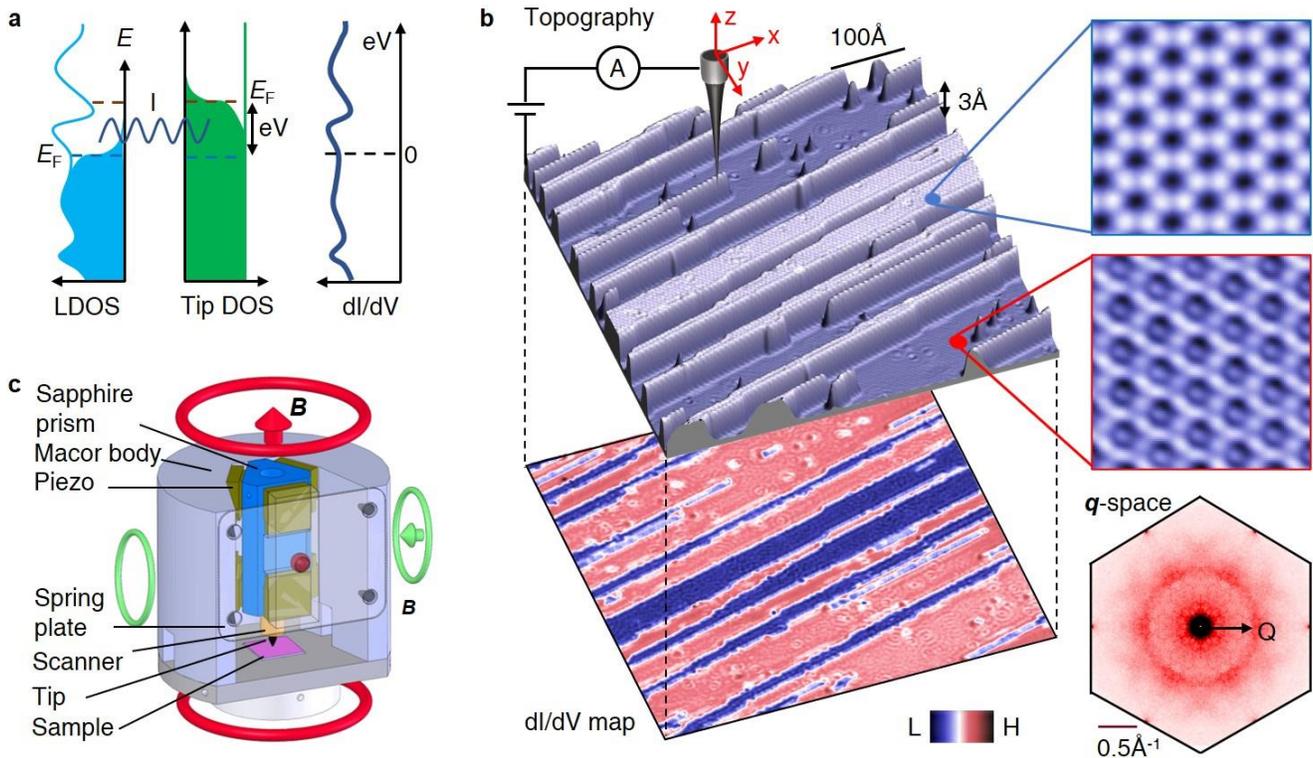

Fig. 1 | **Fundamentals for scanning tunnelling microscopy (STM) of topological quantum materials.** **a** | Schematic for the quantum tunnelling principle of STM. A bias voltage (V) is applied to the sample, which effectively changes its Fermi level by energy eV with respect to the Fermi level of the tip. This allows for tuning of the quantum tunnelling current between the sample (left) and tip (middle). The energy derivative of the tunnelling current dI/dV (right) can effectively measure the local density of states (LDOS) of the sample. **b** | Illustration of microscopy with STM. The schematic shows STM probe of a topological quantum material (CoSn) with complex surface morphology. The two top right panels display the magnified topographic images, showing a honeycomb lattice of $Sn_2$ and a kagome lattice of $Co_3Sn$. The lower left image shows the dI/dV map around the Fermi energy simultaneously obtained for the same area, where defect induced Friedel oscillations are more clearly observed on the kagome surface. The lower right image shows the Fourier transform of the dI/dV map on the kagome lattice, which is the essential quasi-particle interference data for this material. **c** | Schematic of an STM head with a vector magnetic field. In this case, a rigid STM can robustly control the scanning and tunnelling between the tip and sample under the field perturbation. Panel **b** adapted from REF.[32], Springer Nature Limited.

| Technique | Scanning tunnelling microscopy | Angle-resolved photoemission spectroscopy | Magneto-transport | Density functional theory | Tight-binding model |
|---|---|---|---|---|---|
| **Key parameter** | Local density of states | Spectral function | Conductivity tensor | Density functionals | Wavefunction |
| **Variables** | Location; energy; magnetic field; magnetic or nonmagnetic tip; gating; temperature. | Momentum; energy; photon wavelength; photon polarization; spin polarization; temperature. | Electrical field; magnetic field; Landau level sequence; pressure; temperature. | Crystal structures; Elements. | Hopping strength; internal and external fields and interactions. |
| **Unique aspects** | Probing the scattering geometry; atomic spatial resolution; sub-meV energy resolution; tunable magnetic field; Landau quantization; edge state; detecting zero modes. | Band crossing; spin/orbital-momentum locking and texture; probing both 2D surface and 3D bulk band structures. | Fermi surface geometry; electron mobility; chiral anomaly; anomalous Hall and Hall quantization; Phase transitions;. | Band structures and wavefunction; topological index; bulk-boundary correspondence. | Analytical elaboration on the concept of topology and its measurable consequence. |
| **Connection with(in) scanning tunnelling microscopy** | Correspondence between single defects (vortices) and local density of states; correspondence between (bulk) energy gap and edge states. | Correspondence to momentum integrated photoemission signal; energy gaps; charge/spin/orbital texture; inter/intra-band structure scattering. | Effective quasi-particle dispersion and Fermi surface geometry; magnetic field response; effective mass, Fermi velocity, and Fermi length. | Surface dependent density of states and topography; quasi-particle interference; orbital magnetism. | Analytical evaluation of the symmetry-breaking response to topological band structure. |
| **Limitation** | Momentum resolution; requiring atomically flat surfaces; probing surface states and surface-projected bulk states; thermal smearing from tip. | Spatial resolution; requiring fresh and flat surface; occupied states; energy resolution; no magnetic field. | No energy, spatial, momentum or spin resolution; extrinsic scattering mechanism can contribute to the signal. | Strongly correlated systems; phase transitions; surface reconstruction, and disorder effect. | Model-dependent explanation; considering the full symmetry and ingredients in the experiments. |

Table 1 | **Comparison of different techniques in probing topological matter.**

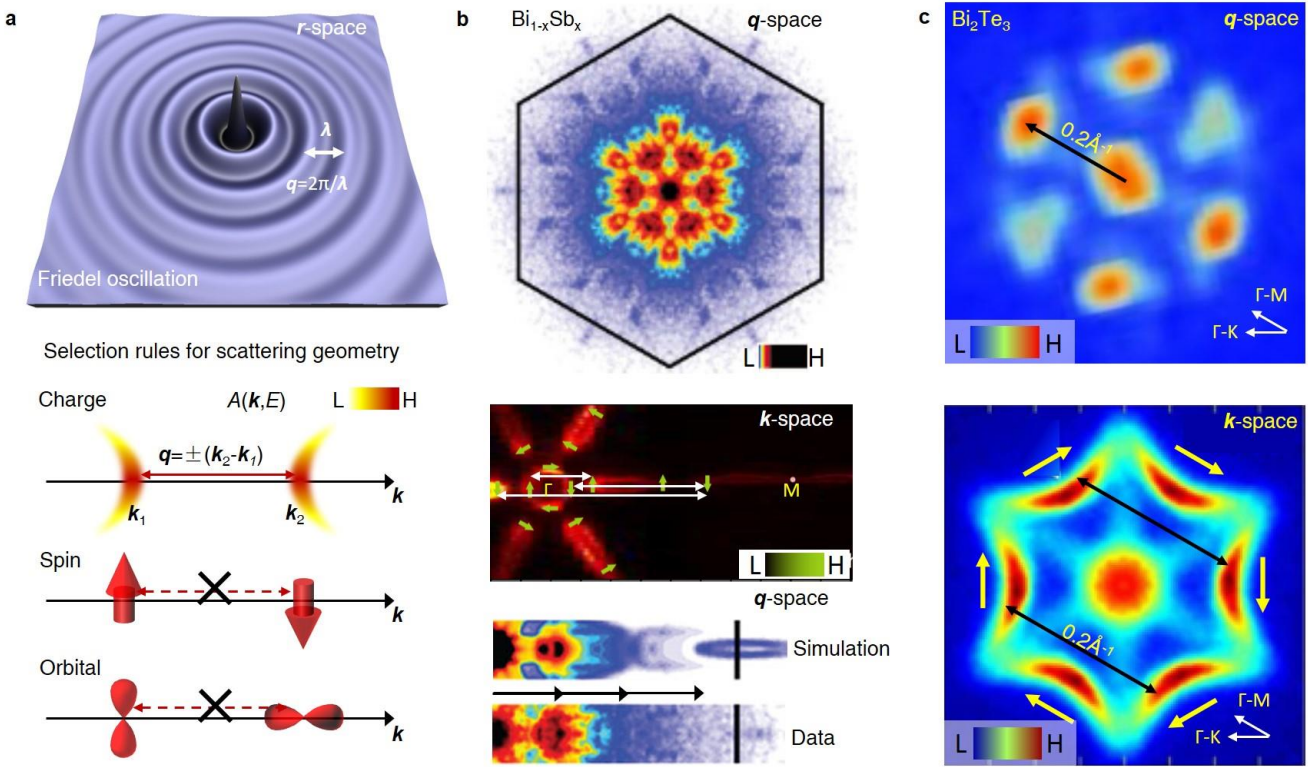

Fig. 2 | **Quasi-particle interference (QPI) method to probe topological scattering geometry.** **a** | Space-momentum nature of the QPI signal. The top panel shows Friedel oscillation observed around an impurity on a metal surface. The $q$ vector in the QPI data corresponds to the period of the oscillation signal in real space. Lower panel illustrates the selection roles for quasi-particle scattering geometry. In momentum space, $q$ wavevector links two different momenta ($k_1$ and $k_2$) in the band structure at a given energy. Three factors in the band structure will affect the QPI intensity at a $q$ vector. Charge: the spectral function intensity at different momenta can vary, and the QPI intensity is proportional to the intensity of the spectral function at $k_1$ and $k_2$. Spin: the electronic structure can feature a momentum dependent spin texture, the QPI intensity is substantially reduced if the spins at $k_1$ and $k_2$ are reversed, especially for nonmagnetic defects assisted QPI. Orbital: the electronic structure can feature a momentum dependent orbital texture, and the QPI intensity is substantially reduced if the orbitals at $k_1$ and $k_2$ are orthogonal. **b** | QPI data for topological insulator $Bi_{1-x}Sb_x$ (top panel), Fermi surface data of the same material obtained by photoemission (middle panel), and the comparison between QPI data and simulated QPI signals based on Fermi surface data along the Γ-M direction (lower panels). The yellow arrows in the middle panel mark the spin texture, and the white arrows mark the three possible scattering vectors along the Γ-M direction. **c** | QPI data (top panel) for topological insulator $Bi_2Te_3$ and its correlation with photoemission signal (bottom panel). The yellow arrows in the lower panel illustrate the spin texture, and the black arrows mark the scattering vector along the Γ-M direction that is detected by QPI. Panel **b** (top and bottom) adapted from REF.[53], Springer Nature Limited. Panel **b** (middle) adapted from REF.[54], AAAS. Panel **c** (top) adapted from REF.[55], APS. Panel **c** (bottom) adapted from REF.[58], APS.

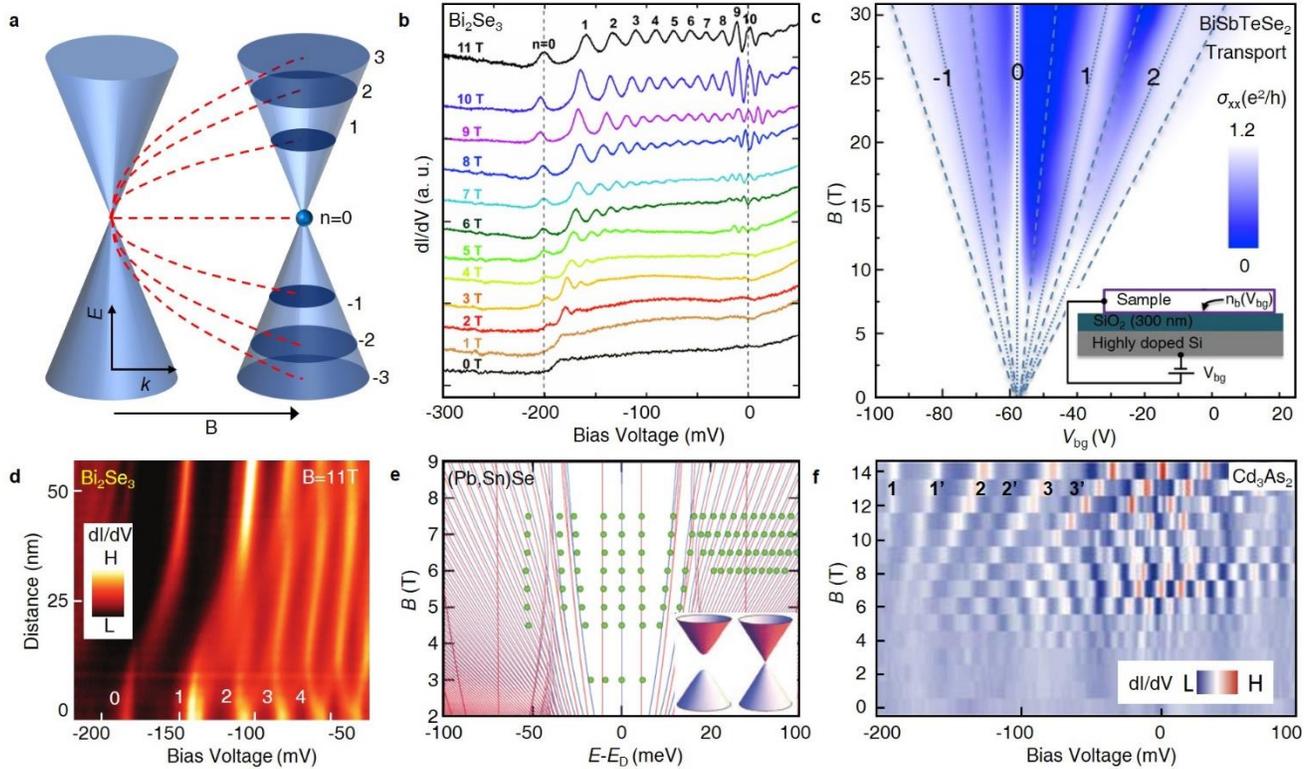

Fig. 3 | **Landau level (LL) quantization method to detect topological fermions. a** | Schematic showing the Landau quantization of Dirac fermions. The field-independent zeroth Landau level, uneven energy distribution of LLs, and nonlinear square root field dependence of the Landau level energy shift are the hallmarks of Dirac dispersion. **b** | Landau level data in the topological insulator $Bi_2Se_3$. The energy of the zeroth Landau level does not change with the field. **c** | Transport Landau fan in the topological insulator $BiSbTeSe_2$ with the inset showing the gating configuration. The gate voltage changes the carrier density that effectively tunes the chemical potential. The energy of zeroth Landau level does not change with the field. **d** | Real space charge potential induced Landau level energy splitting in topological insulator $Bi_2Se_3$. Splitting is most evident for the first Landau level. **e** | Landau level data (green circles) in the topological crystalline insulator (Pb,Sn)Se. There are three field independent Landau levels that can be identified as zeroth LLs, indicating the coexistence of massive (illustrated in the inset as double cones with a gap) and massless Dirac fermions (illustrated in the inset as double cones without a gap). The red and blue lines refer to the theoretical Landau levels from the massive and massless Dirac bands, respectively. **f** | Landau level data in Dirac semimetal $Cd_3As_2$, where two sets of Landau levels are observed. Panel **b** adapted from REF.[88], APS. Panel **c** adapted from REF.[89], Springer Nature Limited. Panel **d** adapted from REF.[91], Springer Nature Limited. Panel **e** adapted from REF.[94], AAAS. Panel **f** adapted from REF.[96], Springer Nature Limited.

Topological kagome magnets refer to a new class of magnetic quantum materials hosting kagome lattice and topological band structure. They include 3-1 materials (example: antiferromagnet $Mn_3Sn$), 1-1 materials (example: paramagnet CoSn), 1-6-6 materials (example: ferrimagnet $TbMn_6Sn_6$), 3-2-2 materials (example: hard ferromagnet $Co_3Sn_2S_2$), and 3-2 materials (example: soft ferromagnet $Fe_3Sn_2$), thus demonstrating a variety of crystal and magnetic structures. They generally feature a 3$d$ transition metal based magnetic kagome lattice with an in-plane lattice constant ~5.5Å. Their 3$d$ electrons dominate the low-energy electronic structure in these quantum materials, thus exhibiting electronic correlation. Crucially, the kagome lattice electrons generally feature Dirac band crossings and flat band, which are the source for nontrivial band topology. Moreover, they all contain the heavy element Sn, which can provide strong spin-orbit coupling to the system. Therefore, this is an ideal system to explore the rich interplay between geometry, correlation, and topology. STM has the unique advantage to resolve the kagome lattice layer of these materials, and perturb the kagome electrons and magnetism with vector magnetic field, uncovering unprecedented topological phases and many-body effects.

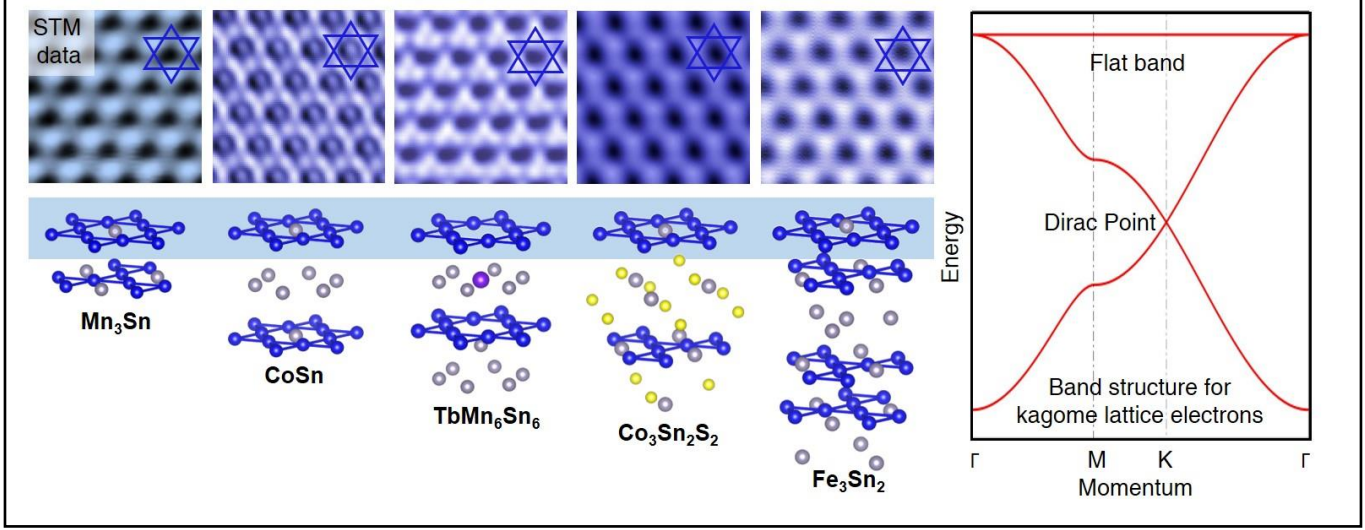

Box 1 | **Topological kagome magnets.**

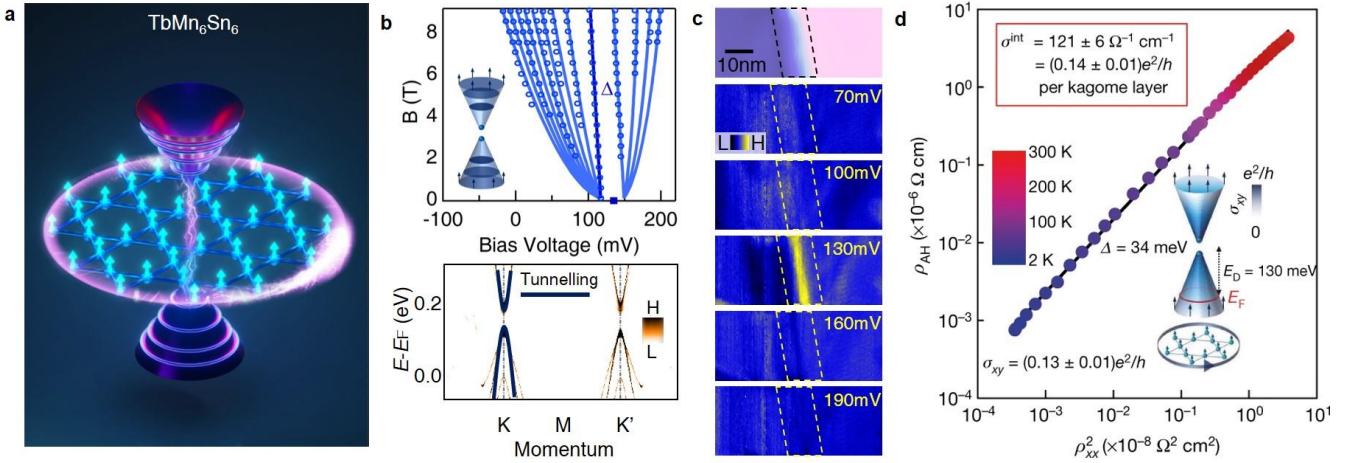

Fig. 4 | **Visualisation of topological bulk-boundary-Berry correspondence. a** | Illustration of the topological correspondence in quantum-limit Chern magnet TbMn$_6$Sn$_6$. In momentum space, spin-polarized Dirac fermions with a Chern energy gap (two separated cones) exhibit Landau quantization. In real space, the spin-orbit-coupled magnetic kagome lattice (spheres with arrows) carries a topological edge state within the Chern energy gap. **b** | Landau level quantization of bulk topological fermions. The upper panel shows the fitting the Landau fan data (open circles) with the spin-polarized and Chern gapped Dirac dispersion (solid lines). Inset: schematic of Landau quantization of Chern gapped Dirac fermions. The lower panel shows the density functional theory of bulk band structure in comparison with the band dispersion obtained from Landau quantization data. **c** | dI/dV maps taken at different energies across a step edge (top). The map taken within the Chern gap energy (130 meV) shows a pronounced topological edge state. **d** | The anomalous Hall resistivity $\rho_{AH}$ plotted against $\rho^2_{xx}$ in a logarithmic scale from 2 K to 300 K. The intrinsic Hall conductance is given by the slope of the line, which amounts to $(0.14 \pm 0.01)e^2/h$ per manganese kagome layer (top inset). The bottom inset illustrates the Berry curvature contribution to the Hall conductivity from Chern gapped Dirac fermions, which is $(0.13 \pm 0.01)e^2/h$ per manganese kagome layer based on the tunnelling data. Panels b-d adapted from REF.[97], Springer Nature Limited.

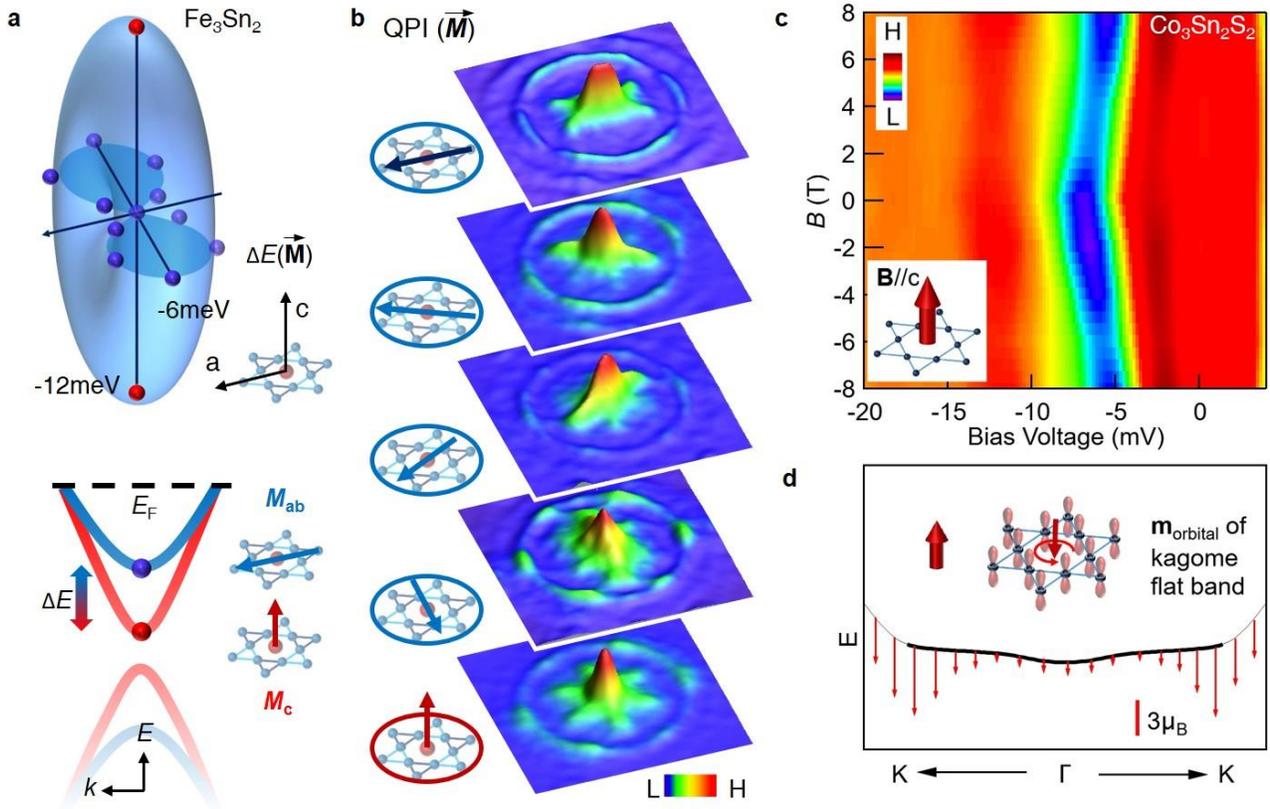

Fig. 5 | **Vector magnetic field control of topological matter. a** | Vector magnetization induced energy-shift of a quantum state in a topological magnet $Fe_3Sn_2$. The top panel shows saturated energy shift ΔE as a function of the direction of the magnetization vector. The light-blue surface shows a 3D illustration of the energy shift ΔE as a function of the magnetization vector, which exhibits a nodal line along the *a*-axis. The lower panel shows the schematic of the magnetization-controlled Dirac gap, with the band bottom of the upper branch corresponding to the shifting state. The red and blue curves illustrate the band structures with in-plane and out-of-plane magnetizations, respectively. **b** | Quasi-particle interference (QPI) patterns of the shifting state as a function of the magnetization direction, which is indicated in the insets with respect to the lattice. The topmost QPI pattern shows the spontaneous nematicity along the *a*-axis. Magnetization along other directions can alter, and thus control, the electronic scattering symmetry. **c** | Vector magnetization control of a flat band state in a topological magnet $Co_3Sn_2S_2$. The inset illustrates that field is applied perpendicular to the kagome lattice plane. **d** | Orbital magnetism for the flat band from density functional theory. The inset illustrates the large negative orbital magnetism of the flat band. Panel **a** and **b** adapted from REF.[80], Springer Nature Limited. Panel **c** and **d** adapted from REF.[113], Springer Nature Limited.

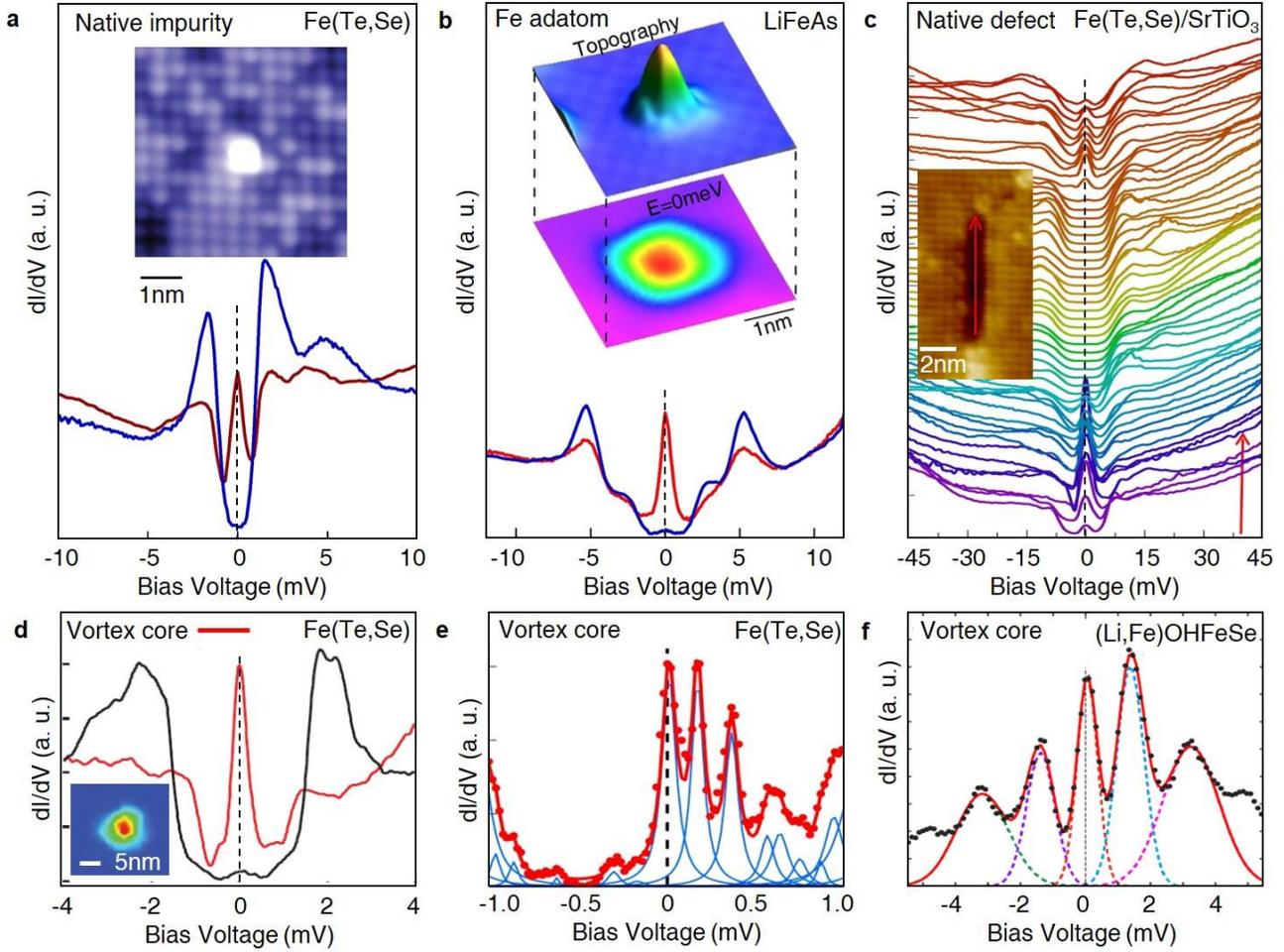

Fig. 6 | **Searching for localized topological zero modes. a** | Zero-energy state bounded to a native interstitial Fe impurity in Fe(Te,Se). The inset shows the topography of the impurity. **b** | Zero-energy state bounded to a Fe adatom deposited on LiFeAs. The inset shows the topography of the impurity and the associated zero-energy dI/dV map. **c** | Zero-energy states bounded at the ends of a line-Te/Se vacancy in Fe(Te,Se)/SrTiO$_3$ film. The inset shows the topography of the line-defect. **d** | Zero-energy state bounded to a vortex core in Fe(Te,Se). The inset shows the dI/dV image of the vortex core. **e** | Zero-energy state bounded to a vortex core in Fe(Te,Se) measured with ultra-high energy resolution. **f** | Zero-energy state bounded to a vortex core in (Li,Fe)OHFeSe. Panel **a** adapted from REF.[38], Springer Nature Limited. Panel **b** adapted from REF.[156], APS. Panel **c** adapted from REF.[159], Springer Nature Limited. Panel **d** adapted from REF.[150], AAAS. Panel **e** adapted from REF.[153], Springer Nature Limited. Panel **f** adapted from REF.[151], APS.

**Acknowledgements**
We acknowledge P. W. Anderson, D. A. Huse, F. D. M. Haldane, N. P. Ong, E. Lieb for discussions on quantum magnets, spin liquid and superconductivity. We acknowledge Abhay Pasupathy, Roland Wiesendanger, Ilija Zeljkovic, Takeshi Kondo, Suyang Xu, Sanfeng Wu, Nirmal Ghimire, Pengcheng Dai, Chin-Sen Ting, Lu Li, Yong P. Chen, Hu Miao, Guang Bian, Yan Sun, Nanlin Wang, David Hsieh, Madhab Neupane, Hao Zheng, Chang Liu, Zhenyu Wang, Canli Song, Jingsheng Wen, Ruihua He, Nan Xu, Ziqiang Wang, Titus Neupert, Biao Lian, Guoqing Chang, Ilya Belopolski, Jing Wang, Gang Su, Jiangping Hu, Gang Xu, Zhong-Yi Lu, Songtian S. Zhang, Hanqing Mao, Bianca S. Swidler, Tyler A. Cochran, Lingyuan Kong and Ruizhe Liu for discussions on STM study of topological matter. Work at Princeton University was supported by the Gordon and Betty Moore Foundation (GBMF4547 and GBMF9461; M.Z.H.). The theoretical work and sample characterization are supported by the United States Department of Energy (U.S. DOE) underthe Basic Energy Sciences programme (grant number DOE/BES DE-FG-02-05ER46200; M.Z.H.). The work on topological superconductivity is partly based on support by the U.S. DOE, Office of Science through the Quantum Science Center (QSC), a National Quantum Information Science Research Center at the Oak Ridge National Laboratory. S.H.P. acknowledges support from the Chinese Academy of Sciences, NSFC (grant no. 11227903), BM-STC (grant no. Z191100007219011), the National Key R&D Program of China (grant nos. 2017YFA0302900 and 2017YFA0302903) and the Strategic Priority Research Program (grant nos. XDB28000000, XDB28010000 and XDB28010200).